\documentclass[singlecolumn,notitlepage,pre,floatfix]{revtex4-1}
\usepackage{graphicx}
\usepackage[english]{babel}
\usepackage{amssymb}
\usepackage{amsbsy}
\usepackage{amsmath}
\usepackage{bbm}
\usepackage{xcolor}

\newcommand{\vx}{{\mathbf x}}
\newcommand{\vv}{{\mathbf v}}

\begin{document}
\title{Pore-scale Mixing and the Evolution of Hydrodynamic Dispersion
  in Porous Media}
 \author{Alexandre Puyguiraud, Philippe Gouze, and Marco Dentz}
 \email[E-mail: ]{marco.dentz@csic.es}
 \affiliation{Spanish National Research Council (IDAEA-CSIC), Barcelona, Spain}
 \affiliation{Geoscience Montpellier, CNRS, Universit\'e de Montpellier, Montpellier, France}

\date{\today}

\begin{abstract}
We study the interplay of pore-scale mixing and network-scale advection through heterogeneous porous media,  
and its role for the evolution and asymptotic behavior of hydrodynamic dispersion. In a Lagrangian framework, we identify 
three fundamental mechanisms of pore-scale mixing that determine large
scale particle motion, namely, the smoothing of intra-pore velocity
contrasts, the increase of the tortuosity of particle paths, and the
setting of a maximum time for particle transitions. Based on these mechanisms, we derive a theory
 that predicts anomalous and normal hydrodynamic
 dispersion based on the characteristic pore length, Eulerian
 velocity distribution and P\'eclet number.
\end{abstract}

\maketitle

\noindent
Transport of 
dissolved substances through porous media
is determined by the complexity of the velocity field in the pore
space and diffusive mass transfer within and between pores.
The interplay of diffusive pore-scale mixing and spatial flow variability are key for
the understanding of transport and reaction phenomena in natural and
engineered porous media~\cite[][]{brenner, sahimi2011flow, Valocchi2018} with diverse applications ranging from
groundwater contamination and geological carbon dioxide storage \cite[][]{Niemi2017}, to the design of batteries~\cite[][]{Maggiolo2020} and transport
in brain microcirculation~\cite[][]{Berg2019}. 
%

Therefore, hydrodynamic transport has been the focus of
research over decades in different disciplines~\cite[][]{josselin1958, saffman1959theory,
  Bear1972, brenner, Isichenko1992, sahimi2011flow}. Still, as
outlined in the following, questions of fundamental nature remain concerning both the evolution of hydrodynamic dispersion, and the
dependence of asymptotic hydrodynamic dispersion coefficients on the
P\'eclet number, which compares the diffusion and advection times over a typical pore length. Anomalous dispersion phenomena~\cite[][]{BG1990} 
have been observed in laboratory
experiments~\cite[][]{Moroni2001,lev03,Seymour2004,morales2017,Carrel2018,Souzy2020},
field scale tracer tests ~\cite[][]{gou08a,haggerty1995multiple}, and numerical
simulations in different types of porous medium and rock
structures~\cite[][]{bijeljic2011signature,Bijeljic2013,deAnna2013,icardi2014pore, kang2014,meyer2016pore,Li2018}. 
They cannot be described by a single constant hydrodynamic dispersion coefficient.
The asymptotic concept of hydrodynamic dispersion models particle displacements due to velocity
fluctuations as Brownian motion and thus implicitly assumes that all
particles have access to the full fluctuation spectrum at each moment,
that is, they can be considered as statistically equal. In a porous medium, however,
velocities vary on length scales engraved in the pore structure, and
thus, particle transitions over regions of low velocity can be much
longer than over regions of high velocity. Statistical
equivalence can only be achieved at times larger than the largest
transition time scale. Thus, anomalous behaviors can be traced back to
broad distributions of mass transfer time scales related to wide
spectra of pore-scale flow velocities.

This phenomenology lies at the heart of non-local transport theories
such as multi-trapping and continuous time random walk (CTRW), 
which have been used to model anomalous and intermittent
pore-scale transport behaviors~\cite[][]{bijeljic2011signature,liu2012applicability,deAnna2013,kang2014,
  Dentz2018, puyguiraud2019upscaling, Nissan2019, Souzy2020}. However, current approaches are limited to purely advective transport, 
or need to be constrained by the measurement of particle transition
times. The quantitative relation between pore-scale mixing, network
scale flow and the evolution of hydrodynamic dispersion remains elusive. 
The pioneering works of \citet{josselin1958} and \citet{saffman1959theory} use the concept of particle
transition times to derive expressions for the asymptotic hydrodynamic dispersion coefficients. Still, and in
spite of numerous theoretical and numerical studies~\cite[][]{Koch1985,Scheven2013,Bijeljic2006}, the
dependence of hydrodynamic dispersion on the P\'eclet number remains an
open issue.    

\begin{figure}[t]
\begin{center}
\includegraphics[width=.35\textwidth]{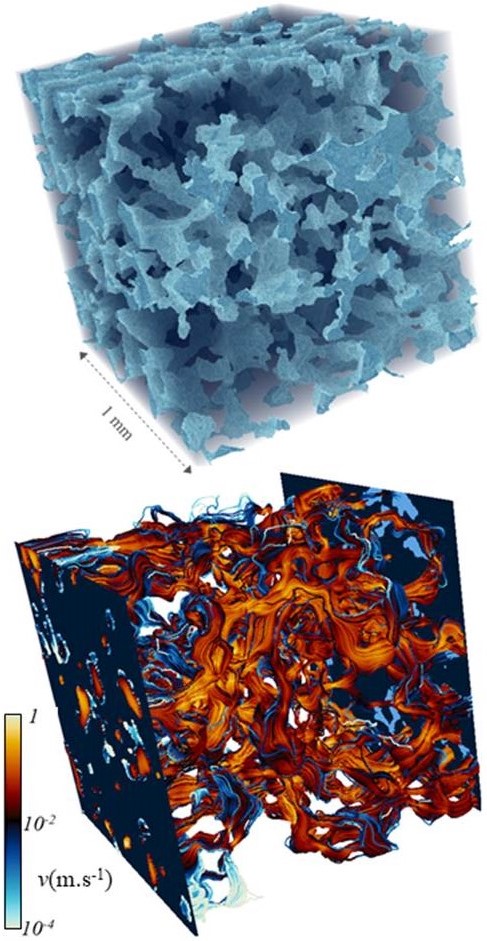}
\end{center}
\caption{Three-dimensional structure of a Berea sandstone
  sample. The top panel highlights the void
  space in shades of blue. The lines in the bottom panel
  show particle paths. The color scheme indicates particle
  speeds from (white-blue-black-orange) low to high.} 
\label{fig:Berea}
\end{figure}

\begin{figure*}
a.\includegraphics[width=.31\textwidth]{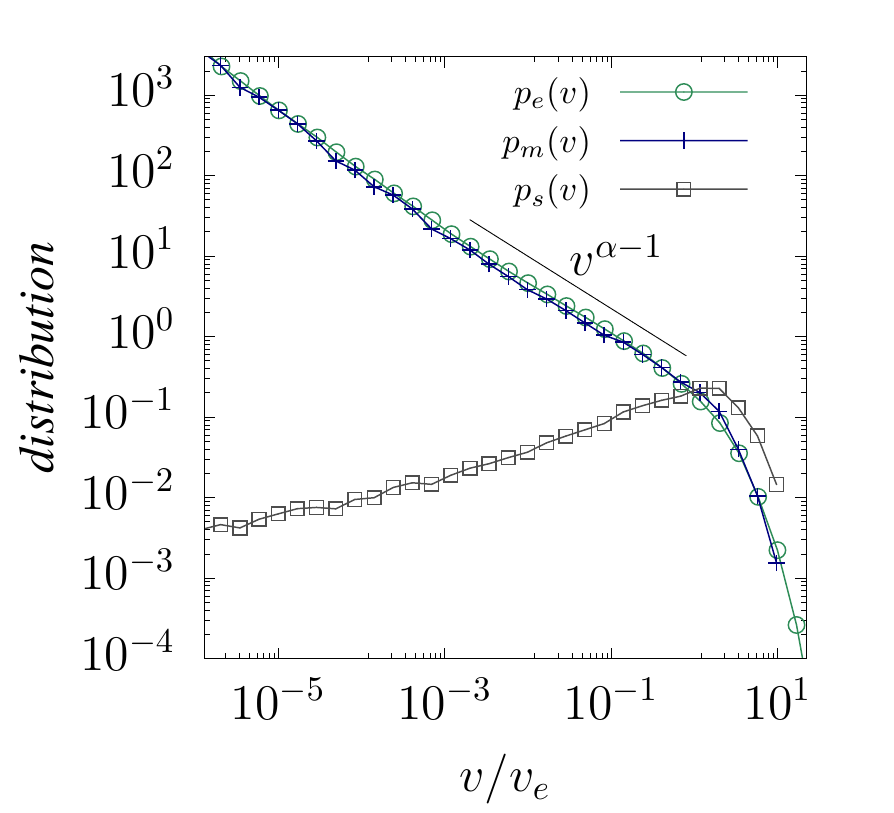}
b.\includegraphics[width=.31\textwidth]{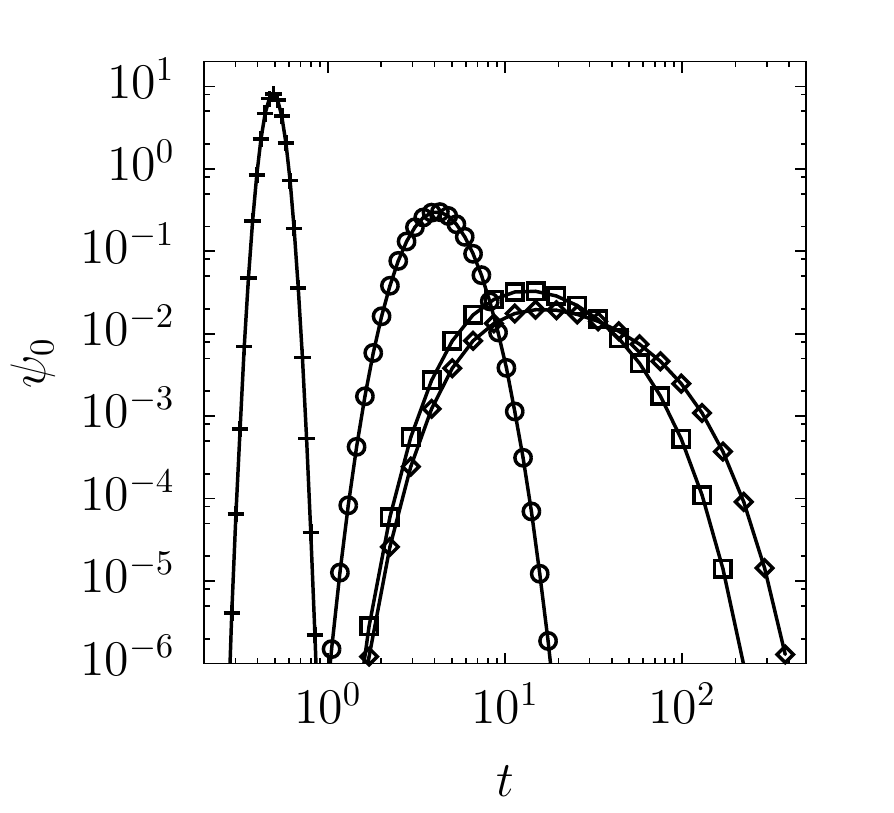}
c.\includegraphics[width=.31\textwidth]{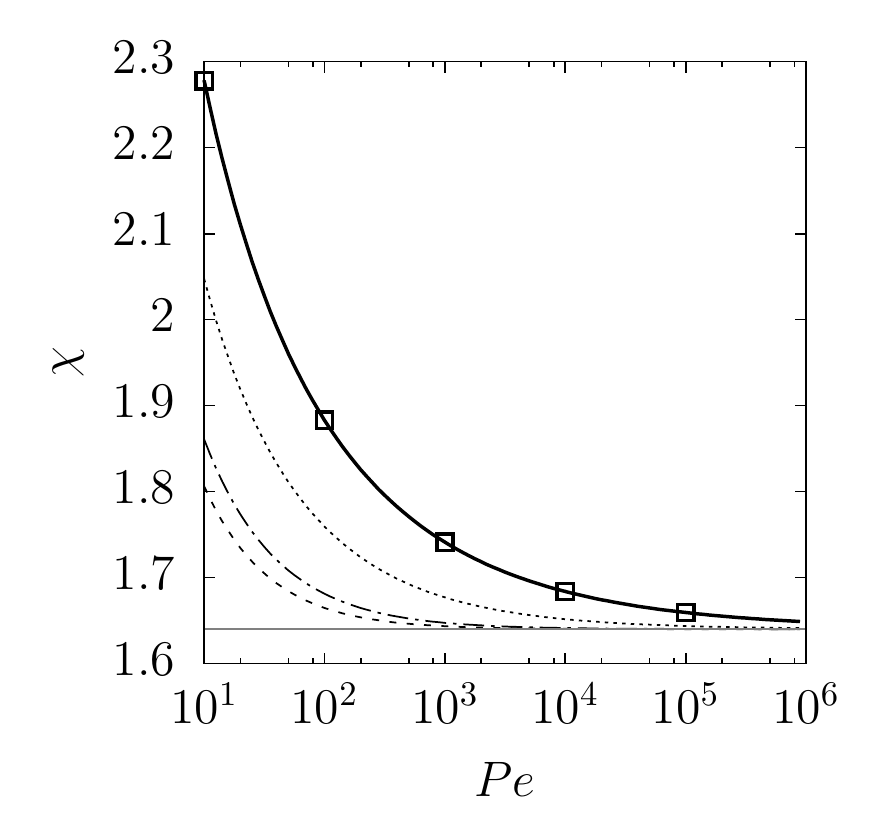}
\caption{(a) PDFs of (circles) Eulerian flow speed $p_e(v)$, (crosses) mean flow speed $p_m(v)$ and (squares) flux weighted mean speed
  distribution $p_s(v)$ normalized by the Eulerian mean speed $\langle
  v_e \rangle$ for the Berea sample of Fig.~\ref{fig:Berea}. The solid line denotes the power-law $v^{\alpha-1}$
  with $\alpha = 0.35$. (b) First passage time distribution
  $\psi_0(t|v)$ for local P\'eclet numbers of (crosses) $Pe_v = 10^2$
  (circles) $10$, (squares) $1$ and (diamonds) $10^{-1}$. (c) Tortuosity $\chi$ versus $Pe$ for Gamma-distributed $p_e(v)$ with exponents (solid line) $\alpha = 0.35$  (dotted) $0.5$, (dash-dotted) $0.8$ and (dashed) $1$. The horizontal line indicates the advective tortuosity $\chi_a$.} 
\label{fig:upscaling}
\end{figure*}

We address these fundamental questions by identifying and quantifying the key mechanisms of pore-scale mixing 
and network-scale flow variability in a stochastic model for the
prediction of hydrodynamic dispersion. We derive a theory that
explains the temporal evolution of dispersion and the
dependence of its asymptotic behavior on the P\'eclet number based on
the Eulerian flow statistics, diffusion and the characteristic
velocity length scale. The theoretical developments are supported and
validated by direct numerical simulations (DNS) of flow and transport
in a $3$-dimensional digitized Berea sandstone sample
(Fig.~\ref{fig:Berea}) obtained using X-Ray
microtomography~\cite[][]{si}. The medium displays strong pore-scale
heterogeneity that gives rise to a broad distribution of flow speeds
illustrated in Figure~\ref{fig:upscaling}a. We consider transport at
different P\'eclet numbers, which are varied by changing the average
flow rate. The molecular diffusion coefficient is set equal to $D_m =
10^{-9}$ m$^2$/s. The P\'eclet number is here defined by $Pe = \langle
v_e \rangle \ell_0/D_m$, where $\ell_0$ is the average pore
length and $\langle v_e \rangle$ the average Eulerian flow speed.


\begin{figure*}[t]
\begin{center}
a.\includegraphics[width=.31\textwidth]{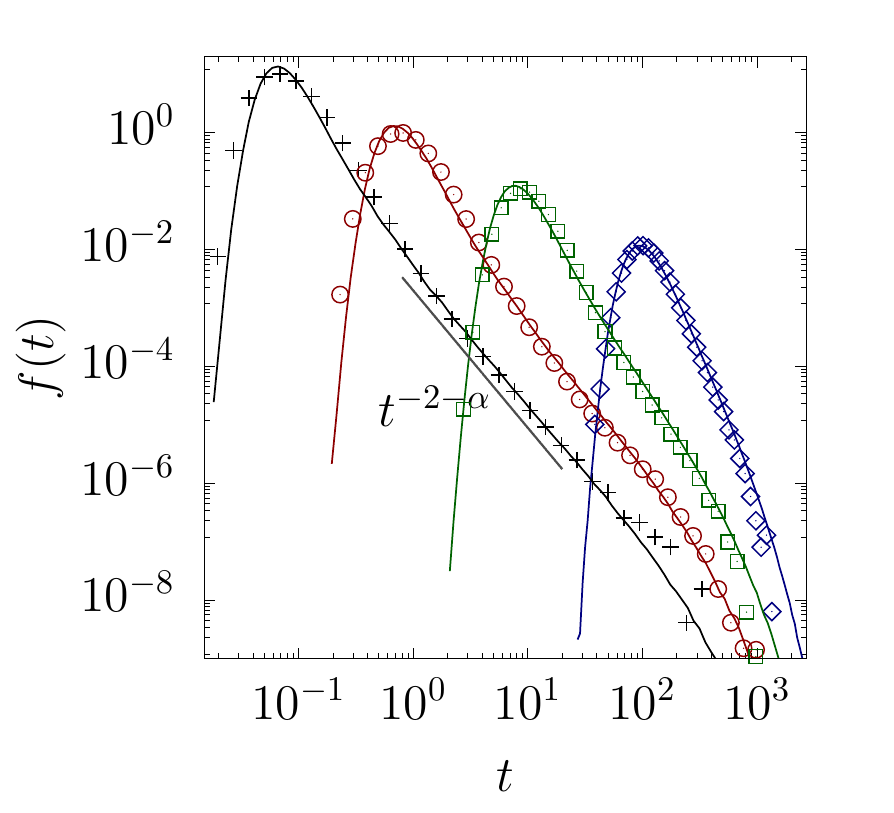}
b.\includegraphics[width=.31\textwidth]{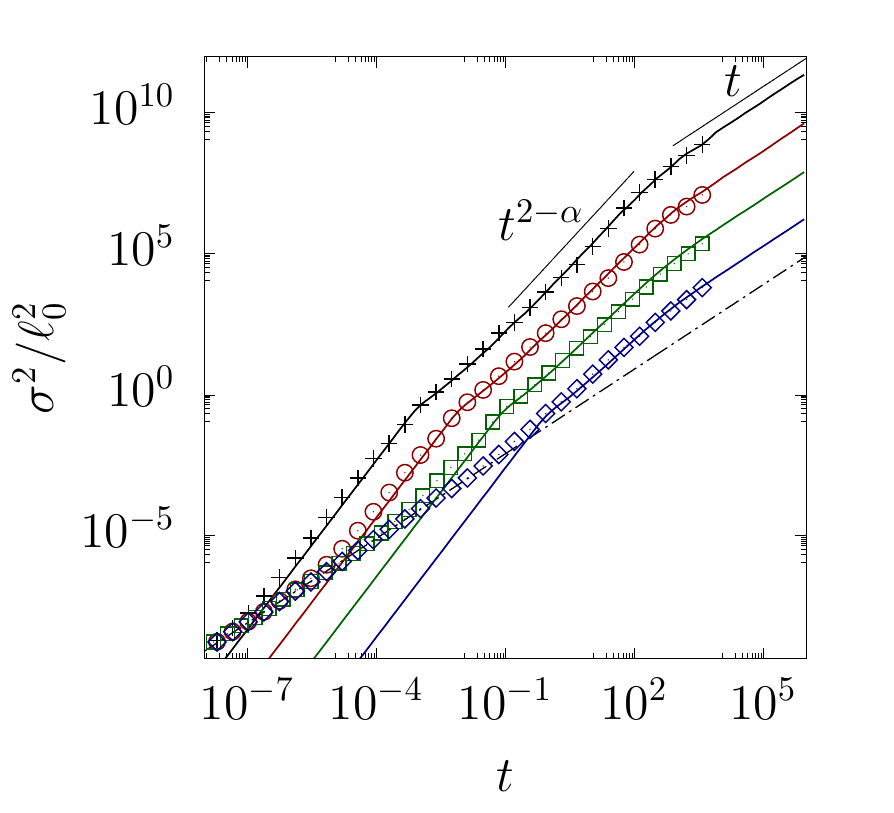}
c.\includegraphics[width=.31\textwidth]{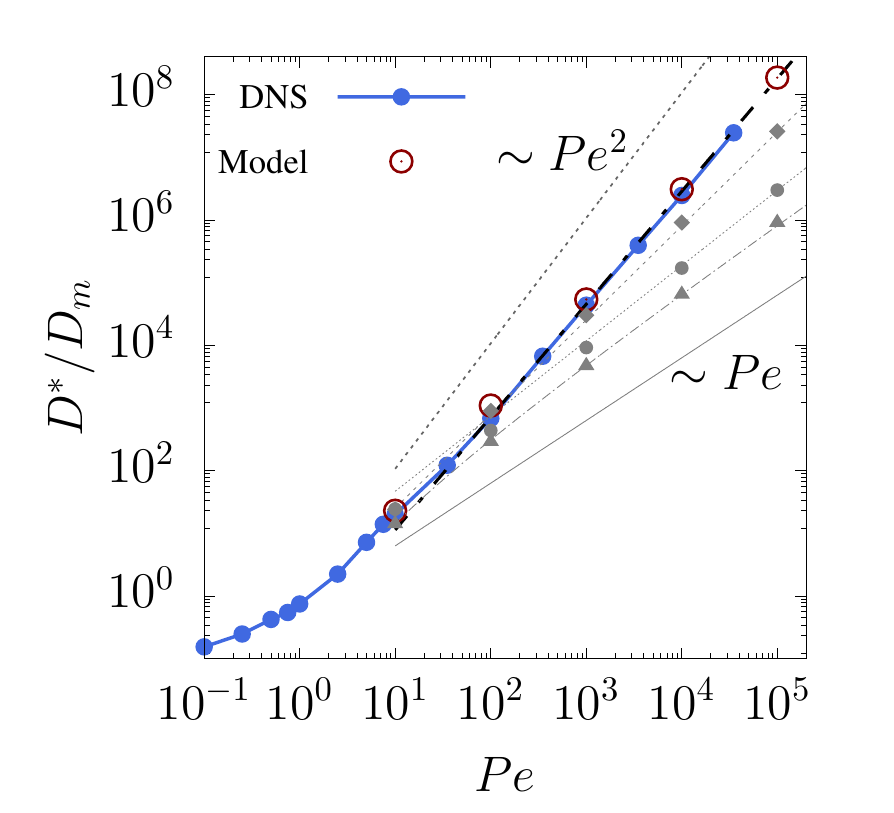}
\end{center}
\caption{(a) First arrival time PDFs at distance $x \approx 32
  \ell_0$ and (b) $\sigma^2(t)$ for (crosses) $Pe = 10^{4}$, (circles) $10^{3}$, (squares) $10^{2}$,
and (rhombs) $10$. (c) Streamwise hydrodynamic dispersion coefficients versus
  P\'eclet number from (full circles) direct numerical simulations, and
  (empty circles) the model predictions. The gray symbols show the
  predictions for Gamma-distributed Eulerian speeds with exponents (top to bottom)
  $\alpha = 0.5,0.8,1$, the corresponding lines show the theoretical
  scalings.}
\label{fig:transport}
\end{figure*}

Figure~\ref{fig:transport} illustrates the impact of flow heterogeneity and diffusion at different $Pe$. 
It shows on the one hand anomalous hydrodynamic dispersion manifest in
heavy-tailed arrival time distributions $f(t)$ and super-diffusive
growth of the longitudinal displacement variance 
$\sigma^2(t)$, and on the other hand cross-overs to asymptotically 
Fickian behaviors (Figs. \ref{fig:transport}a and b). Streamwise dispersion in the asymptotic regime is characterized by the constant hydrodynamic 
dispersion coefficient $D^\ast$, whose non-linear dependence on the P\'eclet number is shown in Figure~\ref{fig:transport}c. 
These features, which result from the complex interplay of flow heterogeneity, diffusion, and geometry are generally
observed in laminar flows through porous media and networks~\cite[][]{bijeljic2011signature}. 

In order to understand the mechanisms that cause these behaviors, 
we consider particle motion along the tortuous
paths in the void space of a porous medium as illustrated in
Figure~\ref{fig:Berea}. The flow speed along streamlines varies over
the correlation scale $\ell_v$ imprinted in the medium geometry and
flow structure~\cite[][]{morales2017,puyguiraudstochastic};  $\ell_v$ is
typically larger than the geometric pore length 
$\ell_0$ due to the tortuosity of the streamlines. 
We model motion along particle pathways through discrete spatial steps along conducts of 
length $\ell_v$ such that subsequent particle speeds $\{v_n\}$ and therefore transition
times $\{\tau_n\}$ along a path can be considered as independent random variables. The distance $s_n$ and transport time $t_n$ of a 
particle along a tortuous pathway are described by the 
stochastic process 
\begin{align}
\label{ctrw}
s_{n+1} = s_n + \ell_v, && t_{n+1} = t_n + \tau_n. 
\end{align}
The travel distance $s(t)$ along a particle path in this
coarse-grained picture is given by  $s(t) = s_{n_t}$, where $n_t =
\max(n|t_n \leq t)$. This picture is equivalent to representing the pore space as a network of 
connected conducts of length $\ell_v$ and average flow speeds $v_m$~\cite[][]{saffman1959theory}, whose intersections correspond 
to the turning points of the process~\eqref{ctrw}. 

For purely advective transport, particles move along streamlines by the local Eulerian
flow speed. Motion along streamlines is projected onto
streamwise motion by advective tortuosity $\chi_a = \langle v_e \rangle/\overline u$,
which is equal to the ratio between the average Eulerian flow speed $\langle v_e \rangle$ and 
average streamwise flow velocity $\overline u$~\cite[][]{Kop1996, Ghanbarian2013,puyguiraud2019upscaling}. 
In the presence of pore-scale diffusion this is different. 
First, during a transition over a conduct, particles sample the flow speeds
across streamlines. Thus, the actual particle speed is different from the
local flow speed along a streamline. Second, diffusion sets a maximum
transition time. In fact, if the local advective particle speed is smaller than $D_m/\ell_v$,
the transition time is diffusion-dominated with a maximum
 of  the order of $\tau_D = \ell_v^2/D_m$. Third, pore-scale mixing increases the effective 
path length and thus tortuosity.  
 
In order to quantify these mechanisms, we first consider the flow speeds by which particles move along the conducts of length $\ell_v$. 
To this end, we assume that flow within a 3-dimensional conduct of length $\ell_v$ can be described by the Poiseuille law, that is, by a 
 parabolic velocity profile. This is a valid approximation because laminar flow in the porous medium is dominated by the drag due to the solid walls. 
If the diffusion time across is smaller than the advection time along the conduct, a particle samples the velocity profile across the conduct and moves effectively with the average conduct velocity. This is the case for narrow conducts associated with low flow velocities. For wide conducts, diffusion removes particles from the low velocities at the grain walls such that the average particle speed is increased towards the mean. 
Thus, particle motion along a conduct is dominated by the mean
speed $v_m$, whose probability density function (PDF) $p_m(v)$ is related to the Eulerian speed PDF $p_e(v)$ by
\begin{align}
\label{pev}
p_e(v) = \int\limits_0^\infty dv' p_m(v) \frac{1}{2 v'} H(2 v'-v).
\end{align}
The assumption of Poiseuille flow gives a uniform speed PDF in the conduct, such that 
the Eulerian PDF can be constructed as the weighted sum of the local uniform PDFs represented by the Heaviside function under the integral. The Eulerian PDF $p_e(v)$ and thus
$p_m(v)$ are determined by volumetric sampling of the local flow
speeds, while the partitioning of particles at turning points is
proportional to the flow rates into the downstream
conducts. Thus, the probability $p_s(v)$ for a particle to choose the speed $v$ at a
turning point is given by the flux-weighted PDF of
mean speeds~\cite[][]{si},
\begin{align}
\label{eq:psOfV}
p_s(v) = \frac{v p_m(v)}{\langle v_m \rangle},
\end{align}
where $\langle v_m \rangle$ is the network averaged mean velocity. 
Figure \ref{fig:upscaling}a displays the Eulerian $p_e(v)$, and $p_m(v)$ for
the Berea sample, and the corresponding Lagrangian
$p_s(v)$. The PDF $p_m(v)$ of mean speeds is very similar to $p_e(v)$ because 
of the power-law behavior at small velocities. 
Equations~\eqref{pev}--\eqref{eq:psOfV}
provide the bridge between Eulerian and Lagrangian flow
characteristics. Current approaches that explore mapping the 
conduct widths to flow speeds via the Poiseuille law and local mass
conservation~\cite[][]{holzner2015intermittent,deannaprf2017,Alim2017,Dentz2018} 
provide promising avenues to ultimately relate $p_e(v)$ to the medium
structure, which, however, is beyond the scope of this paper.  

Next we consider the impact of diffusion on the PDF of
transition times over a single conduct. It is obtained by considering
advective-diffusive transport in a $d = 1$ dimensional domain of length $\ell_v$
characterized by an instantaneous injection at the upstream and an
absorbing boundary condition at the downstream end~\cite[][]{si}. The
transition time PDF $\psi_0(t|v)$ for a single conduct is given by the solute flux
over the downstream boundary. It is sharply peaked around
the advection time $\tau_v = \ell_v/v$ for high local P\'eclet numbers
$Pe_v = v \ell_v/2D$, and broadly distributed with an exponential
cutoff at the diffusion time $\tau_D$ as illustrated in
Figure~\ref{fig:upscaling}b for $Pe_v = 10^2, 10, 1$ and $10^{-1}$. The mean transition time is
given by~\cite[][]{si}
\begin{align}
\label{meant}
\langle \tau|v \rangle = \tau_v \frac{Pe_v - \exp(- Pe_v) \sinh(Pe_v)}{Pe_v}. 
%
\end{align}
%
At high $Pe_v$ it tends to
the local advection time $\tau_v$, at low $Pe_v$ to $\tau_D/2$. The
network-scale PDF $\psi(t)$ of transition times is given in term of $\psi_0(t|v)$ and $p_m(v)$
\begin{align}
\label{psi}
\psi(t) = \int\limits_0^\infty dv \frac{v p_m(v)}{\langle v_m \rangle} \psi_0(t|v). 
\end{align}
For times $t \ll \tau_D$, it can be approximated by $\psi(t) \approx \ell_v^2 t^{-3} p_m(\ell_v/t)$. For times larger than $\tau_D$, $\psi(t)$ decreases to $0$ exponentially fast. Note that this result quantifies particle transition times from first principles, namely pore-scale advection and diffusion, and network scale flow variability.  

So far, we have considered motion along particle paths while our focus
lies on dispersion in streamwise direction, that is, aligned with the mean
pressure gradient. In analogy to purely advective motion, the
increments $\ell_v$ along tortuous particle paths are projected onto
the streamwise increments in terms of tortuosity $\chi$ such that
the particle position $x(t)$ at time $t$ is given by $x(t) =
s(t)/\chi$. In order to determine $\chi$, we note that, under ergodic
flow conditions, the asymptotic mean particle velocity is equal to the
mean flow velocity $\overline u$. This implies that, as particles sample a
representative part of the spatial flow variability they assume
asymptotically the mean flow velocity, which is equal to the Darcy
velocity divided by porosity~\cite[][]{Bear1972}. The stochastic
time-domain random walk (TDRW) model described by
Equations~\eqref{ctrw}--\eqref{psi} yields for the asymptotic average
particle velocity $v_\infty = \ell_v/\chi \langle \tau \rangle \equiv \overline
u$, where $\langle \tau \rangle$ is the average transition time~\cite[][]{Dentz2004}. This
gives for tortuosity  $\chi = \ell_v/\overline u \langle \tau
\rangle$. Using~\eqref{psi} and~\eqref{meant}, we obtain the explicit
expression
\begin{align}
\chi = \chi_a \left[1 - \int\limits_0^\infty dv 
  \frac{1 - \exp(- Pe_v)}{Pe_v} p_m(v)\right]^{-1}. 
\label{tortuosity}
\end{align}
%
The behavior of $\chi$ for the Berea sample is
shown in Figure~\ref{fig:upscaling}c. For $Pe \gg 1$ it tends
to advective tortuosity $\chi_a$ and increases monotonically for decreasing $Pe$.   

We use the derived theory to quantify and elucidate anomalous and normal
hydrodynamic dispersion behaviors. To this end, we focus on speed PDFs that behave as
power-laws for speeds much smaller than the average, $p_m(v) \sim
v^{\alpha-1}$ with $0 < \alpha \leq 1$. Such behaviors have been observed for a wide
range of porous media~\cite[][]{Bijeljic2013,puyguiraudstochastic,Souzy2020}. Note
that $\alpha = 1$ for a channel or tube. For the Berea sample we find
$\alpha \approx 0.35$ as shown in Figure~\ref{fig:upscaling}a. The
corresponding transition time PDF behaves as $\psi(t) \sim (t/\tau_v)^{-2-\alpha}$ for
times $t \ll \tau_D$, and decays exponentially fast for $t \gg
\tau_D$, which is a key feature of the impact of pore-scale mixing. 
To facilitate the analysis, we note that the stochastic
TDRW model described by
Equations~\eqref{ctrw}--\eqref{tortuosity} constitutes a
CTRW so that we can use the CTRW machinery
to derive the hydrodynamic dispersion behaviors predicted by the
theory~\cite[][]{si}. 

We focus on the PDF $f(t)$ of first arrival times
at a control plane perpendicular to the mean flow direction, and the longitudinal displacement variance
$\sigma^2(t) = \langle x(t)^2 \rangle - \langle x(t) \rangle^2$.  
For times $t \ll \tau_D$ particles see only the power-law scaling $\psi(t) \sim t^{-2-\alpha}$ of
the transition time PDF. In this
regime, particle motion is history-dependent because
transition times may be of the order of the observation
time. CTRW based on the generalized central limit theorem predicts power-law scaling as $f(t) \sim
t^{-2-\alpha}$~\cite[][]{berkowitz2006modeling}. The displacement
variance is predicted to scale as $\sigma^2(t) \sim
\overline u \ell_v (t/\tau_v)^{2 - \alpha}$ for $0 < \alpha < 1$ and
as $\sigma^2(t) \sim \overline u \ell_v (t/\tau_v) \ln(t/\tau_v)$
for $\alpha = 1$~\cite[][]{Shlesinger1974,si}. This means, dispersion
is anomalous. For times $t \gg \tau_D$ all particles are able to access
the full spectrum of transition times and become statistically
equal because the memory of the initial velocities is lost. This leads to 
an exponential cut-off in $f(t)$~\cite[][]{Dentz2004}, and linear (Fickian) scaling of $\sigma^2(t)$
 as $ = 2 D^\ast t$. The hydrodynamic
dispersion coefficient $D^\ast$ can be obtained by matching
the preasymptotic and asymptotic expressions for $\sigma^2(t)$ at $t = \tau_D$. This
gives 
%
${D^\ast}/{D_m} \sim  Pe^{2-\alpha}$.    
%
A similar behavior was obtained
by~\citet{bijeljic2011signature} based on empirical transition time PDFs
using CTRW theory~\cite[][]{Dentz2004}. The
theory presented here directly links these behaviors to the
distribution of Eulerian flow speeds. For $\alpha = 1$, we obtain
%
${D^\ast}/{D_m} \sim Pe \ln(Pe)$,
%
which is equivalent to the one derived by~\citet{saffman1959theory}
and~\citet{Koch1985} based on the assumption that
the distribution of flow speeds is flat, which is characteristic of
the linear flow profile at pore walls. 
For $\alpha > 1$, the theory predicts $D^\ast/D_m \sim Pe$. Details
are given in the Supplemental Material~\cite[][]{si}. 

These features of anomalous and asymptotic Fickian dispersion are
illustrated in Figure~\ref{fig:transport} for the Berea sandstone
sample. The power-law scaling and exponential tempering of $f(t)$ are
shown in Figure~\ref{fig:transport}a for different
$Pe$. Figure~\ref{fig:transport}b shows the evolution of
$\sigma^2(t)$. The theory predicts the early
time ballistic behavior, the cross-over to anomalous dispersion for $t
> \tau_v$, and the transition to normal dispersion for
times $t > \tau_D$. The diffusive behavior at very early times $t < D_m/\overline
u^2$ is not resolved by the theory because it does not explicitly
represent (Brownian) particle motion at short times. The impact of
diffusion is accounted for through its effect on particle transition
times, tortuosity and velocity sampling as detailed above. The theory predicts that the dependence of
$D^\ast$ on $Pe$ is constrained between $D^\ast/D_m \sim Pe^2$ and
$D^\ast/D_m \sim Pe$. This is illustrated in Figure~\ref{fig:transport}c, which
shows $D^\ast$ versus $Pe$ obtained from the Berea sample
as well as the theoretical predictions for $\alpha = 0.5, 0.8, 1$
using a Gamma-distributed $p_e(v)$. 

Data from a broad range of experimental and numerical studies of dispersion in a variety of
porous media~\cite[][]{Pfannkuch1963, Bear1972, Bijeljic2004,
  Bijeljic2006,  sahimi2011flow, icardi2014pore} indicate a non-linear
increase of $D^\ast$ with $Pe$. These data are often interpreted jointly by
a single regression~\cite[][]{Bear1972,sahimi2011flow}, which implicitly assumes the existence of a
universal behavior across different types of porous media. The derived
theory indicates that neither the $Pe$-dependence of asymptotic
hydrodynamic dispersion nor its evolution are universal, but depend on 
the flow distribution. On the other hand, the theory shows that
dispersion can be predicted based on the distribution of
flow speeds, which can be applied across a broad range of porous media. 

The derived theory quantifies anomalous and normal hydrodynamic
dispersion from first principles in terms of the characteristic velocity scale, 
the Eulerian flow speed, and pore-scale diffusion. It is valid for $Pe
> 1$ and based on the knowledge of the Eulerian speed distribution. It
is not constrained by transport measurements. The fundamental nature
of the considered flow and transport processes in the conceptual
picture of a network of conducts allows application of the key
elements of the derived theory to transport of dissolved chemicals, 
bacteria and colloids in a wide range of porous media also under
non-Newtonian and multiphase flow conditions. 







\section*{Acknowledgements}
AP and MD gratefully acknowledge the support of the European Research Council (ERC) through the consolidator project MHetScale (617511) and the Spanish Ministry of Science and Innovation through the project HydroPore (PID2019-106887GB-C31).   
The authors gratefully acknowledge the support of the CNRS-PICS
project CROSSCALE. 

\appendix

\section{Direct numerical simulations}
The numerical flow and transport simulations are performed on the 
three-dimensional image of a Berea sandstone sample
 obtained by identifying the connected void
phase and the solid phase by processing a X-Ray microtomography image
(see for example~\citet[][]{Gouze_Melean_2018_WRR_2008} and references
therein)
\subsection{Flow}\label{App:flow}
In the following we summarize the methodology to solve the flow field. 
The binary images of the geometry are composed of $300^3$ regular voxels (cubes) that
represent either void or solid. The mesh used for solving flow is obtain by dividing
each of the image voxels by 3 in each of the direction so that 1 voxel of the image is represented by
27 cubic cells of size $\Delta x = \Delta y =\Delta z = 1.06 \times 10^{-6}$ m. This discretization level is selected such that flow in the smallest
throats are is-represented, see also \citet{gjetvaj2015dual}.  
Thus, the resulting discretization for the regular grid consists of $900^3$ cubic cells.
We prescribe pressure boundary conditions at the inlet and outlet, and
no-slip conditions at the void-solid interfaces and at the remaining
domain boundaries. At the inlet a pressure of $0.1$ Pa in set while it is zero at the outlet.
We then solve the flow with the SIMPLE algorithm~\cite[][]{Patankar1980} implemented in
OpenFOAM \cite[][]{Weller1998}. Note that, in order to
minimize boundary effects, twenty
layers are added at the inlet and outlet~\cite[][]{Guibert2015_MG}. After
convergence, this means, once the residual of the pressure and flow
fields between two consecutive steps is below $10^{-5}$, we
extract the complete velocity field. Velocity values are given at
every interface of the mesh in the normal direction to the
face. More details are given in \citet{gjetvaj2015dual}.
The mean flow speed is $7.78 \times 10^{-7}$ m/s which corresponds to a
Reynolds number of $Re \approx 10^{-5}$, meaning that the flow is laminar
and can be described by the Stokes equation.
The flow fields used for the simulations at different P\'eclet numbers are obtained by multiplying this
flow field by a constant. The corresponding Reynolds numbers are between $Re \approx
10^{-5}$ for $Pe = 1$ and $Re \approx 1$ for $Pe = 10^5$, which is at the
upper limit for which the Stokes assumption is still valid.
\subsection{Random walk particle tracking}\label{App:RWPT}
The random walk particle tracking simulations are based on the Langevin equations
\begin{align}
\frac{d \vx(t)}{d t}= \vv [\vx(t)]+\sqrt{2D}\bm{\xi}(t),
\end{align}
where, $\bm{\xi}=(\xi_1,\xi_2,\xi_3)$ is a Gaussian white noise with zero mean and correlation $\langle \xi_i(t) \xi_j(t') \rangle = \delta(t - t')$.
We can then discretize the Langevin equations as the current position $\vx(t)$ plus an advective and a diffusive component as
\begin{align}
\label{eq:partTrack}
\vx(t+\Delta t)=\vx(t)+\vv[\vx(t)]\Delta t+\sqrt{2D\Delta t}\bm{\zeta}(t).
\end{align}
%
The advective term is based on an extension of the Pollock algorithm
\citep{pol88, Mostaghimi2012, puyguiraudstochastic}. Originally, the Pollock algorithm assumes
a linear variation of velocity within an the mesh cells in each
direction. It is widely used in reservoirs and very high porosity
structures. However, this linear interpolation causes precision errors
in the vicinity of solid surfaces since a linear interpolation is no
longer accurate. This is why \citet{Mostaghimi2012} extended the
methodology by introducing different types of quadratic interpolations
in the voxels that are in contact with the solid phase. This renders this
methodology accurate in low porosity media. This
methodology allows to know analytically the position $\vx(t)$ of a particle for any $t$ and thus permits splitting the trajectory in
small time intervals $\Delta t$. The advective and diffusive
operators are split on this $\Delta t$ basis, allowing for the
computation of the diffusive jumps between advective steps. 

The diffusive jumps are computed following the third term on the right
side of equation \eqref{eq:partTrack} where $\bm{\zeta}=(\zeta_1, \zeta_2,
\zeta_3)$ with $\zeta_i$ being uniform random variables in
$[-\sqrt{3},\sqrt{3}]$ with $\langle \boldsymbol \zeta(t) \rangle =
\mathbf 0$ and $\langle \zeta_i(t) \zeta_j(t) \rangle =
\delta_{ij}$. The central limit theorem guarantees that the sum of the
random motions is Gaussian. Using uniformly generated random variables
rather than Gaussian present two main advantages. The computational
cost is reduced and there is a better control on the maximum
displacement jump that a particle can do. This avoids unexpectedly
large displacement that can jump over solid cells and thus, allows for a moderately large $\Delta t$.

In order to simulate particle displacement over distances larger than the sample size, we reinject the particles at the inlet boundary of the domain once they reach the outlet of the sample. To ensure continuity of the speed series of each particle, we first compute the particle speed $v(\vx_a)$ at position $\vx_a$ at the outlet. Then, we identify the pore area $\mathcal{A}_{v(\vx_a)}$ at the inlet plane where the speed values $v(\vx_b)$ at positions $\vx_b \in \mathcal{A}_{v(\vx_a)}$ fulfill $v(\vx_b) \in [v(\vx_a)-\Delta v, v(\vx_a)+\Delta v]$, where $\Delta v = {v(\vx_a)}/{200}$. The particle is then reinjected in a random location within $\mathcal{A}_{v_L(\vx)}$. This procedure preserves the speed continuity and ensures that no artificial decorrelation is occurring. Besides, in the case of a particle exiting the domain through the inlet, by diffusion, the particle is reinjected at the outlet following a similar procedure.
\section{Speed distributions}
\label{app:VelDistri}
We derive here the distribution of the mean flow speeds and then of the particle speeds that is required in the time-domain random walk approach. 
\subsection{Mean flow speeds}

We conceptualize the porous medium as a network of conducts and
joints. Within each conduct, the volumetrically sampled speed
distribution is uniform, 
\begin{align}
p(v|v_m) = \frac{1}{2 v_m} H(2 v_m - v),
\end{align}
where $v_m$ is the mean speed in the conduct. The Eulerian speed distribution $p_e(v)$
is constructed by integration of $p(v|v_m)$ over all conducts weighted
by the distribution $p_m(v)$ of mean speeds. 
This gives 
\begin{align}
p_e(2 v) = \int\limits_{v}^\infty d v_m p_m(v_m) \frac{1}{2 v_m}.
\end{align}
This implies that the distribution of mean pore speeds can be obtained from the Eulerian speed PDF as
\begin{align}
p_m(v) = - 2 v \frac{d p_e(2 v)}{d v}. 
\end{align}
%
\subsection{Particle speeds}
In the time-domain random walk, the partitioning of particles at
turning points (the joints) is proportional to the flow rate into the downstream
conducts. In our model, the distribution of Eulerian speeds and
thus mean speeds is obtained through volumetric sampling within
the void space of the porous medium,
\begin{align}
p_m(v) = \frac{1}{V_0} \sum_p V_p \delta(v - v_p). 
\end{align}
The distribution $p_s(v)$ of speeds is weighted by the flow rate of the
conducts, which means
\begin{align}
p_s(v) = \frac{1}{\sum_p A_p v_p} \sum_p A_p v_p \delta(v - v_p).
\end{align}
We assume that the length of the conducts is approximately constant
such that $V_p = A_p \ell_0$. Thus, we can write
\begin{align}
p_s(v) = \frac{1}{\sum_p V_p v_p} \sum_p V_p v_p \delta(v - v_p) =
  \frac{v}{\langle v_e \rangle} p_e(v). 
\end{align}
%
\section{Transition time distribution} \label{app:psi}
We first derive the distribution of transition times for a single
conduct, and then the compound transition time distribution on the
network scale. 

\subsection{Single Conduct}

The transition time distribution for a single conduct is obtained from the solution of the following first-passage
problem. We consider an instantaneous injection of tracer at the
upstream turning point at $x = 0$, and an absorbing boundary 
at the downstream turning point at $x = \ell_v$. This means
\begin{align}
\label{app:ade}
\frac{\partial g(x,t)}{\partial t} + v \frac{\partial g(x,t)}{\partial
  x} - D \frac{\partial^2 g(x,t)}{\partial x^2} = 0
\end{align}
with the boundary conditions
\begin{align}
&v g(x,t) - D \frac{g(x,t)}{\partial x} = \delta(t), && x = 0,
\\
&g(x,t) = 0, && x = \ell_v,
\end{align}
and the initial condition $g(x,t = 0) = 0$. The first passage time
distribution over the downstream boundary is given by 
\begin{align}
\psi_0(t|v) = - \left.D \frac{\partial g(x,t)}{\partial x}\right|_{x = \ell_v}.
\end{align}
We solve this first passage problem in
Laplace space. Laplace transform of~\eqref{app:ade} gives 
\begin{align}
\lambda g^\ast(x,\lambda) + v \frac{\partial g^\ast(x,\lambda)}{\partial
  x} - D \frac{\partial^2 g^\ast(x,\lambda)}{\partial x^2} = 0. 
\end{align}
The solution is given by 
\begin{align}
\label{app:solution}
g^\ast(x,\lambda) = A(\lambda) \exp(Pe_v x/\ell_v) \sinh\left[\frac{\ell_v -
  x}{\ell_v} B(\lambda)\right], && B(\lambda) = \sqrt{Pe_v^2 + \lambda
                                   \tau_D},
\end{align}
where we defined $Pe_v = v \ell_v/2D$. The constant $A(\lambda)$ is
determined from the boundary condition at $0$,
\begin{align}
v g^\ast(x,\lambda) - D \frac{g^\ast(x,\lambda)}{\partial x} = 1, && x
                                                                     = 0.
\end{align}
Inserting~\eqref{app:solution} into the latter gives 
\begin{align}
A = \frac{\ell_v/D}{Pe_v \sinh(B) + B \cosh(B) }.
\end{align}
Thus, we obtain $g^\ast(x,\lambda)$
\begin{align}
g^\ast(x,\lambda) = \frac{\ell_v}{D} \frac{\exp(Pe_v x/\ell_v) \sinh\left[\frac{\ell_v -
  x}{\ell_v} B(\lambda)\right]}{Pe_v \sinh(B) + B \cosh(B)}. 
\end{align}
This gives for the first passage time distribution
\begin{align}
\label{app:fptd}
\psi_0^\ast(\lambda|v) = \frac{B\exp(Pe_v)}{Pe_v \sinh(B) + B
  \cosh(B)}. 
\end{align}

\subsubsection{Moments}

The mean travel time is defined by 
\begin{align}
\langle \tau|v \rangle = - \left. \frac{d f^\ast(\lambda)}{d \lambda} \right|_{\lambda = 0}.
\end{align}
We obtain
\begin{align}
\langle \tau|v \rangle 
& = \tau_v \left[ 1 -
  \frac{\exp(- Pe_v) \sinh(Pe_v)}{Pe_v}\right], 
\label{app:mtau}
\end{align}
where we defined $\tau_v = \ell_v/v$. 
The mean squared travel time is defined by 
\begin{align}
\langle \tau^2|v \rangle = \left. \frac{d^2 f^\ast(\lambda)}{d \lambda^2} \right|_{\lambda = 0}.
\end{align}
We obtain
\begin{align}
\langle \tau^2|v \rangle = \tau_v^2 \left(1 - \frac{1}{Pe_v^2} + \frac{\exp(-2 Pe_v)}{Pe_v} \left[3 + \frac{1}{2 Pe_v} + 
\frac{\exp(-2 Pe_v)}{2 Pe_v} \right] \right).  
\label{app:mtau2}
\end{align}
%
\begin{figure}
\includegraphics[width=0.8\textwidth]{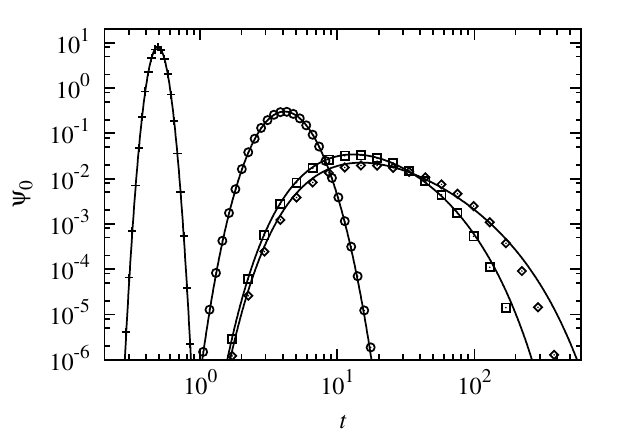}
\caption{First passage time distribution $\psi_0(t|v)$ obtained from numerical inverse Laplace transform
  of~\eqref{app:fptd}  for local P\'eclet numbers of (crosses) $Pe_v = 10^2$ (circles) $10$,
  (squares) $1$ and (diamonds) $10^{-1}$. The solid lines denote the
  corresponding approximations by the
  truncated inverse Gaussian distribution~\eqref{app:IG}. 
\label{fig:psi}}
\end{figure}
\subsubsection{Numerical Approximation}

Numerically, we approximate $\psi_0(t|v)$ by the truncated inverse Gaussian
distribution
\begin{align}
\label{app:IG}
G(t) = \frac{\exp\left[-\frac{\tau_v^{-2}(t-\tau_v)^2}{4
  t/\tau_D} \right]}{t\sqrt{4 \pi t/\tau_D}} \exp(- k t - Pe_v+ \sqrt{Pe_v^2 + k\tau_D}). 
\end{align}
The constant $k$ is chosen such that $G(t)$ has the same mean
transition time as $\psi_0(t|v)$. It is determined as follows. The Laplace
transform of $G(t)$ is given by 
\begin{align}
G^\ast(\lambda) = \exp\left[- \sqrt{Pe_v^2 + (\lambda +k)\tau_D} +
  \sqrt{Pe_v^2 + k\tau_D}\right]. 
\end{align}
The first moment is given by 
\begin{align}
m_G = \frac{\tau_D}{2\sqrt{Pe_v^2 + k\tau_D}} \equiv \langle \tau|v \rangle.  
\end{align}
Thus, we obtain for $k$
\begin{align}
k \tau_D = \frac{\tau_D^2}{4 \langle \tau|v \rangle^2} - Pe_v^2 = Pe_v^2 \left(\frac{\tau_v^2}{\langle \tau|v \rangle^2} -1\right).
\end{align}
Figure~\ref{fig:psi} shows the first passage time distribution and the
approximation by the truncated inverse Gaussian distribution. 
Random numbers are sampled numerically from the truncated inverse Gaussian
distribution by using the algorithm of~\citet{Michael1976} for the
inverse Gaussian random variable in combination with rejection
sampling in order to account for the exponential cutoff.  

\subsection{Network scale}
We first analyze the behavior of the transition time distribution and specifically its behavior for times smaller and larger than $\tau_D$. Then, we consider the behavior of the 
mean and of the mean squared travel time for large P\'eclet numbers. 

The transition time distribution $\psi(t)$ for the network of conducts
is obtained for $\psi_0(t|v)$ and $p_m(v)$ as
\begin{align}
\psi(t) = \int\limits_0^\infty dv \frac{v}{\langle v_m \rangle}
  p_m(v) \psi_0(t|v). 
\end{align}
For $Pe_v \gg 1$, that is for $v \gg D_m/\ell_v$ $\psi_0(t|v)$ is sharply peaked about $\tau_v =
\ell_v/v$, while for $Pe_v \ll 1$, that it $v \ll D_m/\ell_v$ it is $\psi_0(t|v) = \psi_0(t)$ independent of $v$ and a
function of $\tau_D$ only. Thus, we can approximate
\begin{align}
\psi(t) \approx \int\limits_{D_m/\ell_v}^\infty dv \frac{v}{\langle v_m \rangle}
  p_m(v) \delta(t - \ell_v/v) + \int\limits_0^{D_m/\ell_v} dv \frac{v}{\langle v_m \rangle}
  p_m(v)\psi_0(t).
\end{align}
This gives 
\begin{align}
\psi(t) \approx \frac{\ell_v}{\langle v_m \rangle t^3}
  p_m(\ell_v/t) H(\tau_D - t) +  C \psi_0(t), 
\end{align}
with $C$ a constant. Thus, for times $t \ll \tau_D$, the transition
time distribution is dominated by the speed distribution $p_m(v)$, and
for time $t \gg \tau_D$ by the diffusive cut-off
$\tau_D$. 

In the following, we consider speed distributions that behave as the power-law 
\begin{align}
\label{app:pl}
p_m(v) \sim v^{-1-\alpha}
\end{align}
for $v \ll v_0$ with $v_0$ a characteristic velocity, and which decay exponentially fast for $v \gg v_0$. 
For illustration one can think of a Gamma distribution. 
The transition time distribution thus behaves as 
\begin{align}
\psi(t) \sim t^{-2-\alpha}
\end{align}
for $t \ll \tau_D$. 

The mean travel time is given by 
\begin{align}
\langle \tau \rangle &= \int\limits_0^\infty dv \frac{v}{\langle v_m \rangle}
  p_m(v) \langle \tau|v \rangle.
\end{align}
Inserting expression~\eqref{app:mtau} gives
\begin{align}
\langle \tau \rangle &
 = \tau_v \left[ 1 - \int\limits_0^\infty dv
  p_m(v)\frac{\exp(- Pe_v) \sinh(Pe_v)}{Pe_v} \right].
\label{app:mt1}
\end{align}
Next, we rescale the integration variable as $v \to v /v_m$, which gives
\begin{align}
\langle \tau \rangle = \tau_v \left[ 1 - \int\limits_0^\infty dv
  \hat p_m(v) \frac{\exp(- Pe_c v) \sinh(Pe_c v)}{Pe_c v} \right],
\end{align}
where we defined $Pe_c = \langle v_m \rangle \ell_v/2 D_m$ and $\hat p_m(v) = \langle v_m \rangle p_m(\langle v_m \rangle v)$. Note that $Pe_c = Pe \ell_v/\ell_0$.
The leading order behavior of $\langle \tau \rangle$ for $Pe_c \gg 1$ is 
\begin{align}
\label{app:t1}
\langle \tau \rangle = \tau_v + \dots,
\end{align}
where the dots denote contributions of order $Pe$. 

The mean squared travel time is
\begin{align}
\langle \tau^2 \rangle &= \int\limits_0^\infty dv \frac{v}{\langle v_m \rangle} \tau_v^2 p_m(v) F(Pe_v)
= \int\limits_0^\infty dv \frac{\ell_v^2}{v \langle v_m \rangle} p_m(v) F(Pe_v)
=  \tau_v^2 \int\limits_0^\infty dv \frac{\langle v_m \rangle}{v} p_m(v) F(Pe_v),
\label{app:mt2}
\end{align}
where we defined $\langle \tau^2|v \rangle = \tau_v^2 F(Pe_v) $. We 
note that $F(Pe_v)$ behaves for $Pe_v \ll 1$ as $F(Pe_v) = 5
Pe_v^{2}/3$ and is equal to $1$ in the limit $Pe_v \to \infty$. 
Next, we rescale the integration variable as $v \to v /v_m$. This gives  
\begin{align}
\langle \tau^2 \rangle &= \tau_v^2 \int\limits_0^\infty dv \frac{1}{v} \hat p_m(v) F(v Pe_c).
\end{align}
 The mean of $\hat p_m(v)$ is equal to $1$. 
If 
\begin{align}
\int\limits_0^\infty dv v^{-1} \hat p_m(v) < \infty,
\end{align}
this means if $\hat p_m(v)$ goes to $0$ for $v \to 0$, the mean squared transition time behaves as 
\begin{align}
\label{app:t2a2}
\langle \tau^2 \rangle \sim \tau_v^2 
\end{align}
in leading order for $Pe_c \gg 1$. If $\hat p_m(v)$ has an integrable singularity at $0$, this means, if $\hat p_m(v) \sim v^{\alpha -1}$ with $0< \alpha \leq 1$, we rescale the integration variable as $v \to v Pe_c$. This gives  
\begin{align}
\langle \tau^2 \rangle &= \tau_v^2 \int\limits_0^\infty dv \frac{1}{v} \hat p_m(v/Pe_c) F(v).
\end{align}
We set $\hat p_m(v) = v^{\alpha - 1} \varphi(v)$, where $\varphi(v)$ goes toward a constant for $v \to 0$ and decays exponentially fast for $v \to \infty$. For illustration, one may think of a Gamma distribution with mean $1$. Thus, we obtain 
\begin{align}
\langle \tau^2 \rangle &= \tau_v^2 Pe_c^{1 - \alpha} \int\limits_0^\infty dv v^{\alpha-2} \varphi(v /Pe_c) F(v).
\end{align}
For $0 < \alpha < 1$, the leading order behavior in the limit of large $Pe_c \gg 1$ is
\begin{align}
\label{app:t2a01}
\langle \tau^2 \rangle & \sim \tau_v^2 Pe_c^{1 - \alpha}.
\end{align}
For $\alpha = 1$, we can write
\begin{align}
\langle \tau^2 \rangle &= \tau_v^2 \int\limits_0^\infty dv \frac{1}{v} \varphi(v/Pe_c) F(v) \approx \tau_v^2 \int\limits_0^{Pe_v} dv \frac{1}{v} F(v), 
\end{align}
because $\varphi(v/Pe_c)$ sets a cutoff at $Pe_c$. Since $F(v) \to 1$ for $v \gg 1$, we obtain in leading order 
\begin{align}
\label{app:t2aeq2}
\langle \tau^2 \rangle &= \tau_v^2 \ln(Pe_c). 
\end{align}
%

\section{Implementation of the stochastic time-domain random
  walk model} \label{app:PeDepTortu}

The numerical simulations of the derived stochastic particle model is
based on the equations of motion
\begin{align}
x_{n+1} = x_n + \ell_v/\chi, && t_{n+1} = t_n + \tau_n. 
\end{align}
The time increments $\tau_n$ are generated as follows. A speed $v$ is
sampled according to the distribution $p_s(v)$ by inverse
sampling. Then, the transition time is obtained by sampling from
$\psi_0(t|v)$ approximated by the truncated inverse Gaussian
distribution~\eqref{app:IG}. 

The displacement variance is determined as
\begin{align}
\sigma^2(t) = \langle x_{n_t}^2 \rangle - \langle x_{n_t} \rangle^2, 
\end{align}
where $n_t = \max(n|t_n \leq t)$. The first passage time distributions
are determined as 
\begin{align}
t_a =  \sum\limits_{n = 1}^{n_c-1} \tau_n + \tau_{n_c} \frac{x \chi -
  s_n}{\ell_v}, 
\end{align}
where $n_c = \lceil\chi x/\ell_v\rceil$. The interpolation for the
last step is negligible for $n_c \gg 1$. 
\section{Continuous time random walks}
The continuous time random walk framework~\cite[][]{scher1973stochastic,berkowitz2006modeling} gives for the evolution equation of the
particle distribution $p(x,t)$ in the derived stochastic time-domain
random walk model, the following set of equations,
\begin{align}
p(x,t) &= \int\limits_0^t dt' R(x,t') \int\limits_{t-t'}^\infty dt'' \psi(t''),
\\
R(x,t) &= \delta(x)\delta(t) + \int\limits_0^t dt R(x - \ell_v/\chi,t') \psi(t - t').
\end{align}
These equations can be solved for the Fourier-Laplace transform $\tilde p^\ast(k,\lambda)$ of $p(x,t)$ \cite[][]{Shlesinger1974}, which gives
%
\begin{align}
\label{app:pk}
\tilde p^\ast(x,\lambda) = \frac{1}{\lambda} \frac{1 - \psi^\ast(\lambda)}{1 - \exp(i k \ell_v/\chi) \psi^\ast(\lambda)}. 
\end{align}
The Fourier transform is defined here as 
\begin{align}
\tilde \varphi(k) = \int\limits_{-\infty}^\infty dx \exp(i k x) \varphi(x), && \varphi(x) = \int\limits_{-\infty}^\infty \frac{dk}{2 \pi} \exp(-i k x) 
\tilde \varphi(k). 
\end{align}

The first and second displacement moments are defined in terms of $\tilde p^\ast(k,\lambda)$ as
\begin{align}
m^\ast_1(\lambda) &= - i \left.\frac{\partial \tilde p^\ast(k,\lambda)}{\partial k}\right|_{k = 0}
\\
m^\ast_2(\lambda) &= - \left.\frac{\partial^2 \tilde p^\ast(k,\lambda)}{\partial k^2}\right|_{k = 0}. 
\end{align}
Using expression~\eqref{app:pk}, we obtain \cite[][]{Shlesinger1974}
\begin{subequations}
\label{app:moments}
\begin{align}
m_1^\ast(\lambda) &=  v_0 \lambda^{-2} \mathcal K^\ast(\lambda),
\\
m_2^\ast(\lambda) &= 2 D_0 \lambda^{-2} \mathcal K^\ast(\lambda) + 2 v_0^2 \lambda^{-3} \mathcal K^\ast(\lambda)^2, 
\end{align}
\end{subequations}
where we defined 
\begin{align}
\label{app:K}
\mathcal K^\ast(\lambda) = \frac{\langle \tau \rangle \lambda \psi^\ast(\lambda)}{1 - \psi^\ast(\lambda)},
\end{align}
and 
\begin{align}
\label{app:D0}
v_0 = \frac{\ell_v}{\chi \langle \tau \rangle},
&&
D_0  = \frac{\ell_v^2}{2 \chi^2 \langle \tau \rangle}.
\end{align}
%
\subsection{Asymptotic transport}
In order to determine the asymptotic large scale transport properties, we expand the kernel~\eqref{app:K} up to linear order in $\lambda$
\begin{align}
\mathcal K^\ast(\lambda) = 1 + \lambda \mathcal K^\infty, && \mathcal K^\infty = \frac{1}{2} \frac{\langle
  \tau^2 \rangle - 2 \langle \tau\rangle^2}{\langle \tau\rangle}. 
\end{align}
Thus, we obtain for~\eqref{app:moments}
\begin{align}
m_1^\ast(\lambda) &=  v_0 \lambda^{-2} (1 + \lambda \mathcal K^\infty),
\\
m_2^\ast(\lambda) &= 2 D_0 \lambda^{-2} (1 + \lambda \mathcal K^\infty) + 2 v_0^2 \lambda^{-3} (1 + \lambda \mathcal K^\infty)^2. 
\end{align}
Inverse Laplace transform gives 
\begin{subequations}
\label{app:moments2}
\begin{align}
m_1(t) &= v_0 (t + \mathcal K^\infty)
\\
m_2(t) & = 2 D_0 (t + \mathcal K^\infty) +  v_0^2 (t+ \mathcal K^\infty)^2 + 2 v_0^2 \mathcal K^\infty t + {\mathcal K^\infty}^2.
\end{align}
\end{subequations}
The mean velocity and hydrodynamic dispersion coefficient are defined by 
\begin{align}
v_\infty &= \frac{d}{dt} m_1(t)
\\
D^\ast &= \frac{1}{2} \frac{d}{dt} \left[m_2(t) - m_1(t)^2\right]. 
\end{align}
Using~\eqref{app:moments2}, we obtain 
\begin{align}
\label{app:vinfty}
v_\infty &= v_0
\\
D^\ast &= D_0 + v_0^2 \frac{1}{2} \frac{\langle \tau^2 \rangle - 2 \langle \tau\rangle^2}{\langle \tau\rangle} = D_0 \left(1 +  \frac{\langle \tau^2 \rangle - 2 \langle \tau\rangle^2}{\langle \tau\rangle^2}\right) = D_0 \frac{\sigma_\tau^2}{\langle \tau \rangle^2}.
\label{app:Dast}
\end{align}
%
\subsubsection{Tortuosity}
The macroscopic transport velocity is equal to the mean pore velocity $v_\infty = \overline u$~\cite[][]{Bear1972}.  Thus, we obtain from~\eqref{app:vinfty} with \eqref{app:D0} that the tortuosity is given by 
\begin{align}
\chi = \frac{\ell_v}{\overline u \langle \tau \rangle}.
\label{app:chi}
\end{align}
Using~\eqref{app:mt1} gives the explicit expression
\begin{align}
\chi = \chi_a \left[ 1 - \int\limits_0^\infty dv
  p_m(v)\frac{\exp(- Pe_v) \sinh(Pe_v)}{Pe_v} \right]^{-1},
\end{align}
where $\chi_a = \langle v_e \rangle/\overline u$.
\subsubsection{Hydrodynamic dispersion coefficient}
Equation~\eqref{app:chi} together with~\eqref{app:D0} in~\eqref{app:Dast} gives for the hydrodynamic dispersion coefficient
\begin{align}
D^\ast = \overline u^2 \frac{\sigma_\tau^2}{2\langle \tau \rangle}.
\end{align}
The full P\'eclet dependence of $D^\ast$ can be obtained by using expressions~\eqref{app:mt1} and~\eqref{app:mt2} for the mean and mean squared transition times. 

We determine now the leading order behavior of $D^\ast$ for $Pe_c \gg 1$. For $0 < \alpha < 1$, we obtain by using~\eqref{app:t1} and~\eqref{app:t2a01}
\begin{align}
D^\ast \sim \overline u^2 \tau_v Pe_c^{1 - \alpha} = \langle v_m \rangle \ell_v \chi_a^{-2}, Pe_c^{1 - \alpha}
\end{align}
and therefore
\begin{align}
D^\ast/D_m \sim Pe_c^{2 - \alpha}.
\end{align}
Similarly, we obtain for $\alpha = 1$ by using~\eqref{app:t2aeq2}
\begin{align}
D^\ast \sim \overline u^2 \tau_v \ln(Pe_c) = \langle v_m \rangle \ell_v \chi_a^{-2} \ln(Pe_c),
\end{align}
and thus 
\begin{align}
D^\ast/D_m \sim Pe_c \ln(Pe_c). 
\end{align}
For $\alpha > 1$, we obtain by using~\eqref{app:t2a2}
\begin{align}
D^\ast \sim \overline u^2 \tau_v = \langle v_m \rangle \ell_v \chi_a^{-2},
\end{align}
%
and thus 
\begin{align}
D^\ast/D_m \sim Pe_c. 
\end{align}
%
\subsection{Anomalous dispersion}
Anomalous dispersion is measured by the displacement variance $\sigma^2(t) = m_2(t) - m_1(t)$. The first and second displacement moments $m_1(t)$ and $m_2(t)$ are given in Laplace space by~\eqref{app:moments}. 
At times $t \ll \tau_D$, the transition time distribution behaves as $\psi(t) \sim t^{-2 - \alpha}$. For $0 < \alpha < 1$, its Laplace transform can
be expanded as 
\begin{align}
\psi^\ast(\lambda) = 1 - \lambda \langle \tau \rangle + b \lambda^{1 +
  \alpha},
\end{align}
with $b$ a constant. Inserting this expansion into~\eqref{app:K}, we obtain in leading order for $\mathcal
K(\lambda)$ 
\begin{align}
\mathcal K(\lambda) = 1 + b \lambda^{\alpha} /\langle \tau \rangle.  
\end{align}
Thus, we obtain for $m_1^\ast(\lambda)$
\begin{align}
m_1^\ast(\lambda) = v_0 \lambda^{-2} + b \lambda^{\alpha -2}/\langle \tau \rangle 
\end{align}
This implies
\begin{align}
m_1(t) = v_0 t + \dots,
\end{align}
where the dots denote subleading contributions of order
$t^{1-\alpha}$. For the second moment, we obtain
\begin{align}
m_2^\ast(\lambda) = 2 D_0 \lambda^{-2} + 2 v_0^2 \lambda^{-3} + 2 b
  v_0^2 \lambda^{\alpha-3} /\langle \tau \rangle,
\end{align}
Inverse Laplace transform gives
\begin{align}
m_2(t) = 2 D_0 t + v_0^2 t^2 + 2 v_0^2 b t^{2 - \alpha}/\langle \tau \rangle \Gamma(1-\alpha). 
\end{align}
Thus, we obtain for the displacement variance
\begin{align}
\sigma^2(t) \sim t^{2 - \alpha}. 
\end{align}

For $\alpha = 1$, the Laplace transform of $\psi(t)$ can be expanded
as 
\begin{align}
\psi^\ast(\lambda) = 1 - \langle \tau \rangle \lambda - c \lambda^2 \ln(\lambda),
\end{align}
with $c$ a constant. Inserting this expansion into~\eqref{app:K}, we obtain in leading order for $\mathcal
K(\lambda)$ 
\begin{align}
\mathcal K(\lambda) = 1 - c \lambda \ln(\lambda) /\langle \tau \rangle.  
\end{align}
Thus, we obtain for $m_1^\ast(\lambda)$
\begin{align}
m_1^\ast(\lambda) = v_0 \lambda^{-2} + \dots
\end{align}
This implies
\begin{align}
m_1(t) = v_0 t + \dots,
\end{align}
where the dots denote subleading contributions. For the second moment, we obtain
\begin{align}
m_2^\ast(\lambda) = 2 D_0 \lambda^{-2} + 2 v_0^2 \lambda^{-3} - 2 v_0^2 c
  \lambda^{-2} \ln(\lambda) /\langle \tau \rangle.
\end{align}
Inverse Laplace transform gives
\begin{align}
m_2(t) = 2 D_0 t + v_0^2 t^2 +  2 v_0^2 c' t \ln(t)/\langle \tau \rangle,
\end{align}
with $c'$ a constant. Thus, we obtain for the displacement variance
\begin{align}
\sigma^2(t) \sim  t \ln(t).
\end{align}
%

%
\end{document}